# Comparison between the effect of activated waters on lentil seed germination using various plasma reactors and hydrogen injection system

S. Mansory[1], M. Bahreini[1*], S. H. Tadi[2]

[1] School of physics, Iran University of Science and Technology, Tehran, Iran

[2] Laser and plasma research institute, Shahid Beheshti University, Tehran, Iran

([* ]Corresponding author email: *m_bahreini@iust.ac.ir*)

(Dated: )

As a threat to meeting the global demand for food created by the continued growth of the population, different methods are being applied for enhancing seed germination and plant growth. This study investigates the effect of hydrogen-riched water (HRW), plasma-activated water (PAW) and their combination on the seed germination of lentils. For activating water, arc discharge reactors are generated under atmospheric pressure in the air. Simulations were also performed to simulate electric field and potential, fluid flow, and arc plasma. Using optical emission spectroscopy, species were evaluated in plasma columns. Raman spectra and physicochemical properties of water were investigated. On day 3 after treatment, the fraction and length of germinated seeds were evaluated. During germination, treated water significantly increased germination parameters such as final germination percentage, mean germination time, germination index, coefficient of germination velocity and germination rate index. It can, therefore, be concluded that seed germination can be increased using PAW and hydrogenated PAW combined.

## I. Introduction

A threat to meeting global demand for food has been created by the continued growth of the population all over the world. Various techniques, such as different fertilizers, pesticides, water treatment, heat disinfection, etc., are being applied to increase crop yields[1–4]. However, these methods also have a certain number of problems which include adverse impacts on health crops, eventuate pollution after the treatments and adding toxicity to produce various disease challenges for animals and humans as well as economic impossibility etc.[2,5–7]

As we strive towards finding new ways of promoting sustainable and economically viable solutions for enhancing seed germination and plant growth, one option we are currently exploring is plasma activated water (PAW). PAW is created by transferring various reactive oxygen-nitrogen species (RONS) from the plasma phase to water. This process results in the formation of a range of primary species such as atomic oxygen, single oxygen, superoxide, ozone, hydroxyl radicals, and atomic nitrogen, which are then followed by secondary species such as hydrogen peroxide, peroxy-nitrite, nitric oxide, nitrates, and nitrite ions. These secondary species react with water to form PAW, potentially promoting healthy plant growth[8,9]. The dissolution of plasma species in water causes the formation of short-lived species (O, · OH, ONOOH, and NO, etc.) and long-lived species ($NO_3^-$, $NO_2^-$, $H_2O_2$, and dissolved $O_3$, etc.)[10]. The existence of these reactive radicals and species in PAW makes it a propitious product that can be employed in countless applications, including eliminating or reducing the spread of pathogens[11], improving soil fertility[12], seed germination and plant growth[9], etc.

PAW has been shown to affect seed germination and plant growth with prior reports from various researchers[9,13,14]. Rathore et al.[13] water activated with a pencil plasma jet (PPJ) improves the germination rate, viability index, and mean germination time of peas (Pisum sativum L.). Sivachandiran et al.[9] have reported the positive effect of air plasma and PAW with a DBD reactor on germination rate and plant growth in 3 different seed species, including radish, tomatoes and sweet pepper. Terebun et al.[14] describes the significant improvement of PAW produced in an atmospheric pressure gliding arc reactor for the germination of beetroot (Beta vulgaris) and carrot (Daucus carota) seeds.

Additionally, a series of events in the growth and development of plants have been attributed to H2[15,16]. The role and basic mechanisms for H2 in promoting seed germination were investigated by some researchers, such as Huang et al.[17]. In recent years, $H_2$ is an entirely new antioxidant in animals and plants. As a new beneficial gaseous molecule, $H_2$ can respond to physiological processes[18–20]. In addition, $H_2$ responds to certain abiotic stress factors such as salinity [21], osmotic stresses[22], heavy metals[23], high light stress[24] and temperatures[25]. For instance, an increase in the antioxidant system that counteracts the overproduction of ROS and lipid peroxidation has led to increased salt tolerance of Arabidopsis through Hydrogen-Rich Water (HRW) [21].

Furthermore, the cultivation of food legumes, which provides more than 1 billion people with a key source of digestible carbohydrate polymers, could provide opportunities for use in new plant protein-based foods and animal feeding stuff. Seeds of grain legumes are very popular as a source of digestible carbohydrates, compared to animal-based protein, which is environmentally costly. The value of food, handling characteristics and gustatory qualities can be provided by carbohydrate polymers[26]. The ancient crop, which is produced in more than 70 countries, is Lentils (Lens culinaris Medik.)[27]. The seeds come in a variety of colors and patterns and have a lens-like shape[28]. Large green lentils are sold mainly in European markets and parts of the Middle East and South America[29]. Lentil seeds offer a significant source of dietary fiber, complex carbohydrates, iron, zinc, and vitamin B complex. These nutritional components are essential for maintaining a healthy diet and providing the body with the necessary nutrients for optimal performance[30]. They also contain specific phenolic compounds, which give them a high antioxidant activity compared to other legume species[31].

Different methods of spectroscopy such as optical emission and Raman spectroscopy can provide important information on chemical agents. Raman spectroscopy is an efficient way of obtaining chemical information by means of the distribution of electromagnetic radiation through molecular vibrations in matter[32]. Due to the reduction of or elimination of limitations on Raman Spectroscopy, it is beginning to be applied more and more for qualitative analysis[33–35].

In previous work, finite element simulation investigated two-dimensional dielectric barrier discharge and arc discharge in a bubble inside water[36]. The present study uses an HRW generator and two different atmospheric pressure air plasma generators to produce PAW. In order to gain a better understanding of the electrical and fluidic parameters of plasma, simulation was performed to explore electric fields and potentials as well as fluid flow and plasma arcs. Species were evaluated in plasma columns by optical emission spectroscopy and Raman spectra of water were investigated. Multifarious produced waters were exposed to green Lentils (Lens culinaris Medik.) seeds to study the various phenotypic, growth parameters, etc.

## II. Material and method
### A. Setups

The schematic of the water activating setups illustrated in FIG.1. An output power of 15 kV at 55 kHz frequency and a maximum intensity of 130W is provided by the high voltage power supply used for plasma generation. In this research, PAW production systems are based on arc discharge plasma. A glass tube, an air pump, a rubber pipe and two cylindrical electrodes with a diameter of 1 mm shall be part of the 1st system. The discharge shall take place between the electrodes in a glass tube connected to an air pump. Through a plastic tube, the species formed in the plasma are injected into the water (FIG.1. a). A needle electrode with a shape of 0.7 mm, placed at a height of approximately 3.5 cm from the water surface, and a flat electrode placed in the water, make the other configuration. In this system, the plasma plume is formed between the needle electrode and the water and is always in contact with the water (FIG.1. b).

The HRW generator is composed of two titanium meshes called electrodes, a porous layer and plastic tubes designed to remove oxygen from the chamber (FIG.1. c). The electrode is connected to a 12volt DC power source, with an electrical power of 13 watts.

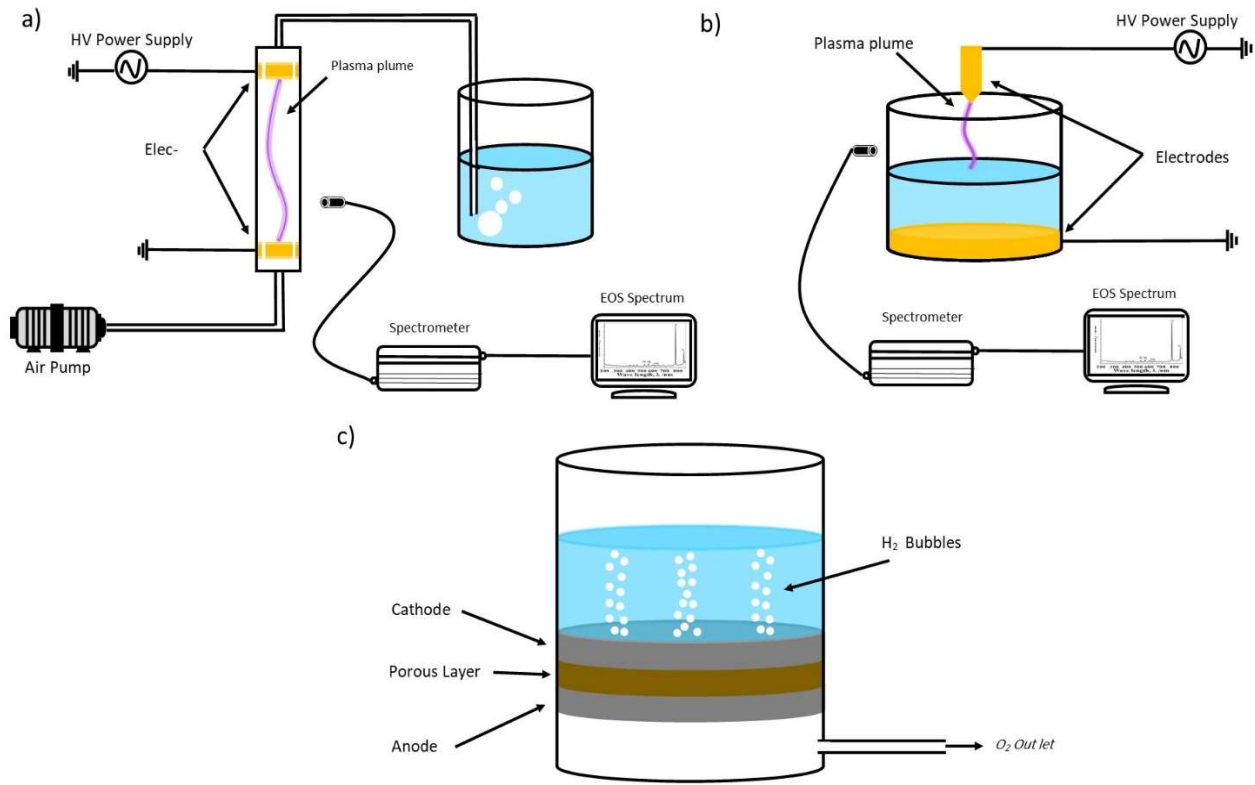

FIG. 1. Schematic of the water activating setups; a) PAW1 (injection of electrically arc discharged air into water); b) PAW2 (arc discharge on the water surface); and c) HRW generator

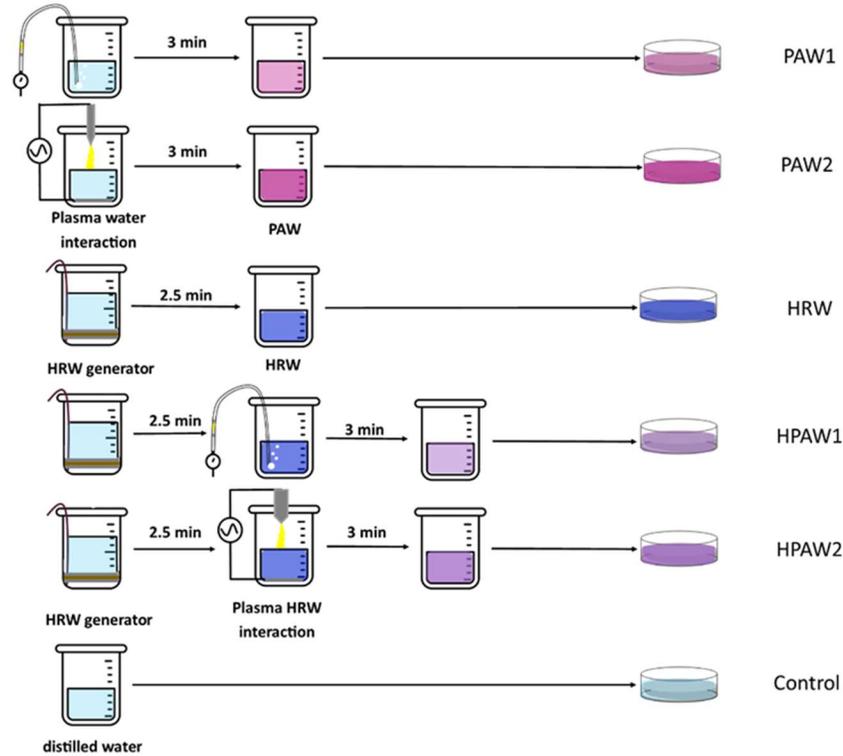

FIG. 2. Schematic of production of various water utilized in this study

**B. Water activation producer**

50 ml of distilled water (control) was taken in 100 ml of a glass beaker. This water was exposed to plasma with two different PAW generators, namely injection of electrically arc discharged air into water (PAW1) and arc discharge on the water surface (PAW2) with a 3 min exposure time. 50 ml of distilled water was taken in the HRW generator and enriched for 2.5 min. Then, activated by PAW generators for 3 min. FIG.2 shows schematic of production of various water that was utilized in this study.

**C. Preparation of lentil seeds**

Green dried lentil seeds (L. culinaris) were purchased from the local market. A total number of 750 seeds (6 groups'× 5 replicate× 25 seeds) were divided into 6 groups named control, PAW1, PAW2, HPAW1, HPAW2, and HRW. Each group was further subdivided into 5 sets of 25 seeds. Each set (25 seeds) of the group soaked in 7 ml of sample (Control, PAW1, PAW2, HPAW1, HPAW2, and HRW) for 72 h (24 °C, 26% Relative humidity).

**D. Simulation**

The Electrostatic Module within Comsol Multiphysics which enables the simulation of electric fields and charged particles and understanding the electrical parameters of plasma, such as electric potential distribution, electric field strength, and charge density as well as, the Laminar Flow Module which simulates the flow of fluid within the plasma reactor are used in this study. It considers the laminar flow regime, where fluid particles move in smooth layers or streamlines without significant mixing. This module helps to investigate plasma fluidic parameters, such as velocity profiles, pressure distribution, and flow patterns.

Two different 2D geometries were designed for two plasma reactors, as illustrated in Figure 3. This figure illustrates the indirect arc reactor in FIG.1.a and a direct arc reactor in FIG.1.b. For simulating the electric field inside a plasma reactor, it is possible to define material properties, boundary conditions, and applied potentials by using the Electrostatic Module. Also, the Laminar Flow Module specifies fluid properties, inlet/outlet conditions, and flow behavior within the reactor.

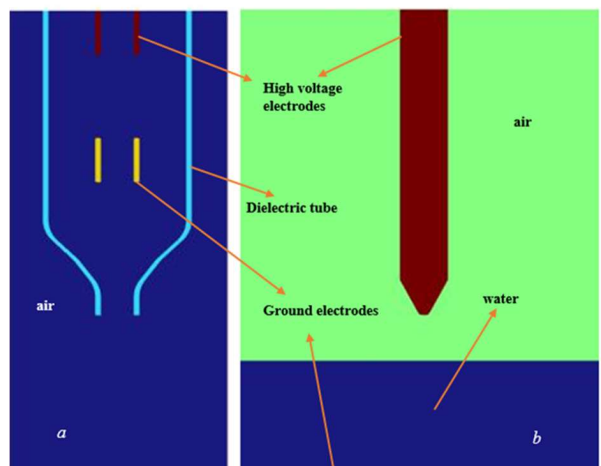

FIG. 3. The schematic geometry of two indirect arc reactors (a) and a direct arc reactor(b).

**E. Plasma chemistry (principles and mechanisms of plasma for water interaction)**

More than a dozen studies focused on the discovery or suggested active particles and their mechanisms of action. A total of 79 gas phase species, 83 water phase species, 1680 gas phase reactions and 448 plasma reactions are included in the complete reaction mechanism[37], which can be summarized as follows:

*1. Chemical properties of gas-phase plasma without water*

Chemical reactions taking place during discharge in atmospheric air (its primary components, typically nitrogen and oxygen) resulting in the formation of ROS and RNS.

$$O_2 + e \rightarrow O^+ + O + 2e \quad (1)$$

$$O_2 + e \rightarrow O^- + O \quad (2)$$

$$O + O_2 \rightarrow O_3 \quad (3)$$

$$O_3 + NO \rightarrow NO_2 + O_2 \quad (4)$$

$$N + O_2 \rightarrow NO + O \quad (5)$$

$$O + N_2 \rightarrow NO + N \quad (6)$$

$$O + NO_2 \rightarrow NO + O_2 \quad (7)$$

$$2 NO + O_2 \rightarrow 2 NO_2 \quad (8)$$

$$3 NO_2 + H_2O \rightarrow 2 HNO_3 + NO \quad (9)$$

$$NO + NO \rightarrow N_2 + O_2 \quad (10)$$

$$NO_2 + h\nu \rightarrow NO + O^\bullet \quad (11)$$

$$NO_3 + h\nu \rightarrow NO + O_2 \quad (12)$$

$$NO_2 + NO_3 \rightarrow N_2O_5 \quad (13)$$

$$3NO_2^- + 3H^+ \rightarrow 2 NO + NO_3^- + H_3O^+ \quad (14)$$

*2. Chemical properties of gas-phase plasma interaction with water*

Plasma reacts to water, triggering a range of kinetic chemical reactions and giving rise to the existence of an Aqueous reactive species. Different processes, e.g. the transfer of gases to liquids and a chemical reaction between gas types and liquid molecules are performed when PAW is generated at an interface layer[38]. The interaction of water with the plasma results in non-equilibrium dissociation of water molecules, leading to the formation of formation of short-living species such as hydroxyl ions (OH⁻) and hydrated (solvated) electrons. A more stable species, including super oxides, ozone, $H_2O_2$, is formed by subsequent rapid reactions between hydroxyl ions and hydrated(solvated) electrons[39]. Apart from ROS, PAW also contain nitric ($HNO_3$) and nitrous acid (HNO$_2$) and low-level transient RNS, e.g. peroxynitrous acid/peroxynitrite and nitrogen dioxide radicals. Nitrogen oxide, which then reacts with water to form acids, is formed in the presence of air when nitrogen and oxygen are separated from the gaseous phase. The reactions are as follows[37,40]:

$$H_2O + e \rightarrow OH^\bullet + H^\bullet + e^- \qquad (15)$$

$$H_2O + e \rightarrow H^+ + OH\bullet + 2e^- \qquad (16)$$

$$H_2O + e \rightarrow H^\bullet + O^\bullet + H^\bullet + e^- \qquad (17)$$

$$NO_2 + OH \rightarrow HNO_3 \qquad (18)$$

$$H_2O_2 + h\nu \rightarrow OH^\bullet + OH^\bullet \qquad (19)$$

$$H_2O_2 + H^+ + NO_2^- \rightarrow ONOOH + H_2O \qquad (20)$$

$$OH^\bullet + OH^\bullet \rightarrow H_2O_2 \qquad (21)$$

$$NO + OH^\bullet \rightarrow HNO_2 \qquad (22)$$

$$HNO_2 + OH^\bullet \rightarrow NO_2 + H_2O \qquad (23)$$

$$NO_2 + NO_3 \rightarrow N_2O_5 \qquad (24)$$

$$2\,NO_2 + H_2O \rightarrow NO_2^- + NO_3^- + 2H^+ \qquad (25)$$

$$OH + NO_2 \rightarrow [O=N-OOH] \rightarrow O=N-OO^- + H^+ \qquad (26)$$

### 3. Chemical properties of gas-phase plasma interaction with HRW

With the presence of hydrogen in HRW, other reactions can also occur. The formula of these reactions is given below[37,37,40].

$$e^- + H_2 \rightarrow e^- + H_2 \qquad (27)$$

$$e^- + H_2 \rightarrow e^- + H + H \qquad (28)$$

$$e^- + H_2 \rightarrow 2e^- + H_2^+ \qquad (29)$$

$$H + O + H_2 \rightarrow OH + H_2 \qquad (30)$$

$$H + O + H_2O \rightarrow OH + H_2O \qquad (31)$$

$$H + O_2 + H_2 \rightarrow HO_2 + H_2^+ \qquad (32)$$

$$H + O_2 + O_2 \rightarrow HO_2 + O_2 \qquad (33)$$

$$H + O_2 + H_2O \rightarrow HO_2 + H_2O \qquad (34)$$

$$H + OH + H_2 \rightarrow H_2O + H_2 \qquad (35)$$

$$H + OH + O_2 \rightarrow H_2O + O_2 \qquad (36)$$

$$H + O_3 \rightarrow OH + O_2 \qquad (37)$$

$$H + O_3 \rightarrow O + HO_2 \qquad (38)$$

$$H + HO_2 \rightarrow H_2O + O \qquad (39)$$

$$H + HO_2 \rightarrow O_2 + H_2 \qquad (40)$$

$$H + HO_2 \rightarrow 2OH \qquad (41)$$

### F. Methods of Calculations of Germination Parameters

Seed germination was monitored daily. Germination data comparison between groups has been done using: Final germination percentage, Mean germination time, Germination index, Coefficient of velocity of germination and Germination rate index. The methodology of calculations of the mentioned parameters is as follows [41]:

1. Final germination percentage (FGP):

$$FGP = Final\ number\ of\ seeds\ germinated\ in\ seed\ lot \times 100$$

2. Mean germination time (MGT):

$$MGT = \frac{\sum F.x}{\sum F}$$ ; where f=Seeds germinated on day x

3. Coefficient of velocity of germination (CVG):

$$CVG = \frac{N1+N2+\ldots+Ni}{100 \times (N1T\ \ldots+NiTi)}$$ ; where N is the number of seeds germinated every day and T is the number of days from seeding corresponding to N.

4. Germination rate index (GRI)

$GRI = \frac{G1}{1} + \frac{G2}{2} + \cdots + \frac{Gi}{i}$; where G1 is the germination percentage on day 1, G2 is the germination parentage at day 2; and so on.

5. Germination index (GI):

$GI = (3 \times N1) + (2 \times N2) + (1 \times N3)$ ; where N1, N2 and N3 is the number of germinated seeds on the first, second and third day and the multipliers (*e.g.*, 1, 2, and 3) are weights given to the days of the germination.

## III. Results and Discussion
### A. Simulations

FIG.4 shows the distribution of the electric potential and the field lines for the electric field. Additionally, in Figure 5 we can see the 2D lines of electric field for direct and indirect arc reactors and the differences between them as well.

The intensity of the electric field must be at least 10 MV/m in order for to an electric discharge occurs at atmospheric pressure. A comparison is shown in FIG.5 of the intensity of the electric field generated by two reactor geometries; it is shown that in both of these geometries, an electric discharge can be generated to the required limit value.

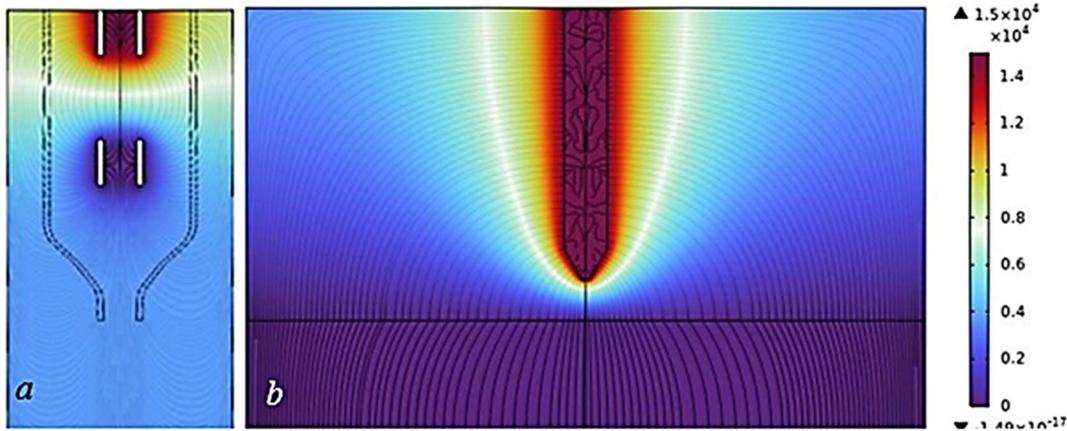

FIG. 4. 2D surface of potential and electric field lines for indirect arc reactor (a) and direct arc reactor (b).

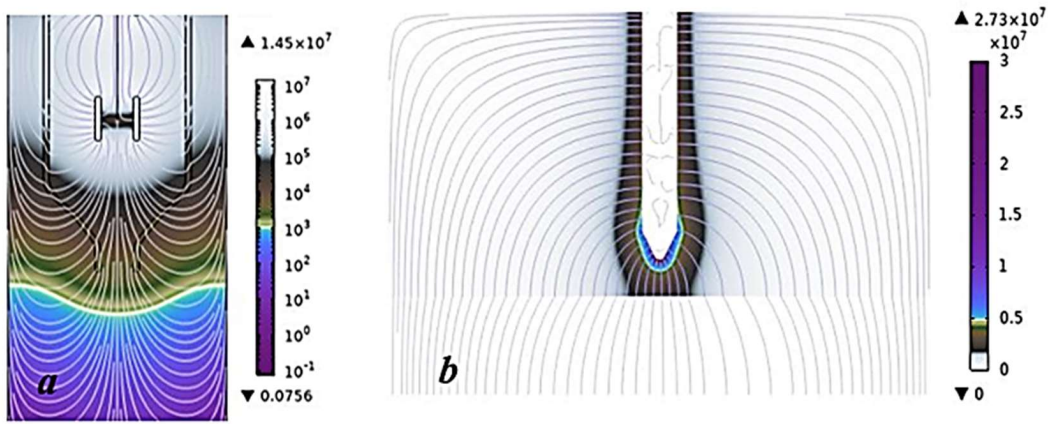

FIG. 5. 2D electric field lines for indirect arc reactor (a) and direct arc reactor (b).

In order to investigate the effects of electric discharge on reactor behavior and fluid dynamics, the arc discharge was modeled by applying a space charge, which is shown in FIG.6. As shown in FIG.6.a, the arc discharge pattern is unstable as a result of fluid flow. In contrast, in direct arc discharge as illustrated in FIG.6.b, it is symmetrical and intense.

Despite the fact that the direct discharge reactor does not have an inlet gas as can be seen in FIG.7, the movement of the charge particles within the arc initiates a fluid flow both within and around the arc. As a result of fluid flow, active species produced in plasma are transferred to water, then the fluid speed at the water surface is an important parameter. Also due to the movement of fluid within the electric arc area, water enters the discharge area, leading to the formation of active hydrogen species.

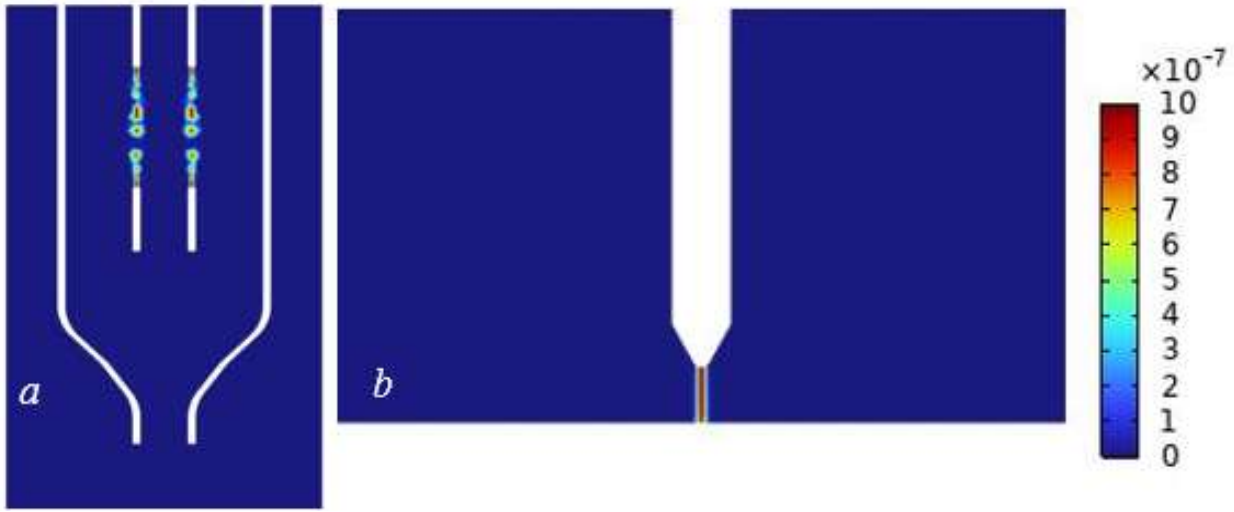

FIG. 6. 2D Arc and space charge pattern for indirect arc reactor (a) and direct arc reactor (b).

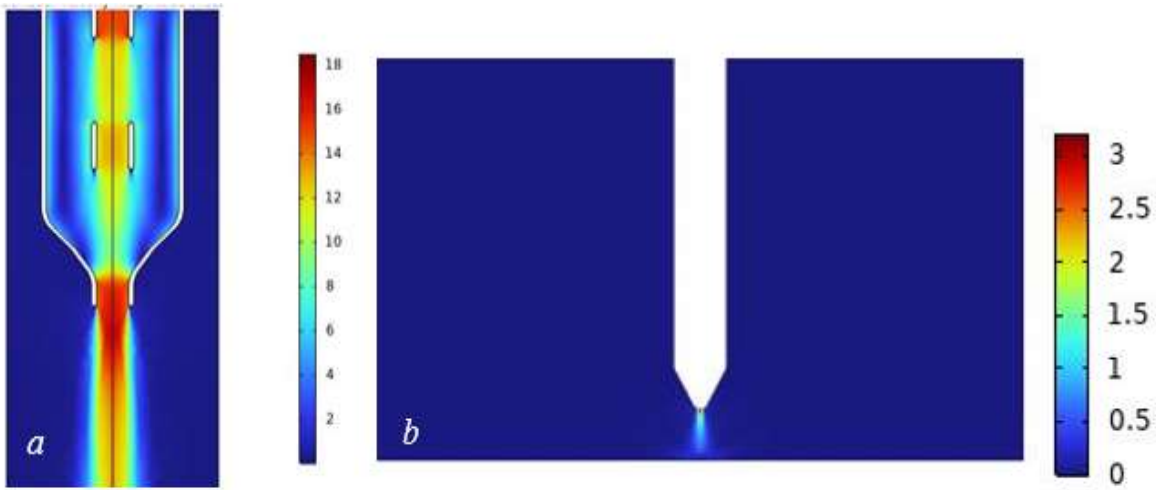

FIG. 7. 2D velocity pattern for indirect arc reactor (a) and direct arc reactor (b).

## B. Optical Emission Spectra

FIG.8 illustrates the optical emission spectra of GAD and Glow discharge electrolysis at wavelengths ranging from 100 to 200 nm. An optical emission spectrometer, Technooran (Noora200), was used to identify the reactive species present in the generated discharge.

At wavelengths ranging from 200 nm to 270 nm, the low-intensity excited nitric oxide (NOγ-band) dominates the spectra[42]. The existence of the hydroxyl radical (OH) is indicated by the signal at 309 nm. At wavelengths ranging from 315 nm to 380 nm, several high and low-intensity nitrogen molecules (N2_the second positive system) were also discovered[43]. In the spectrum of

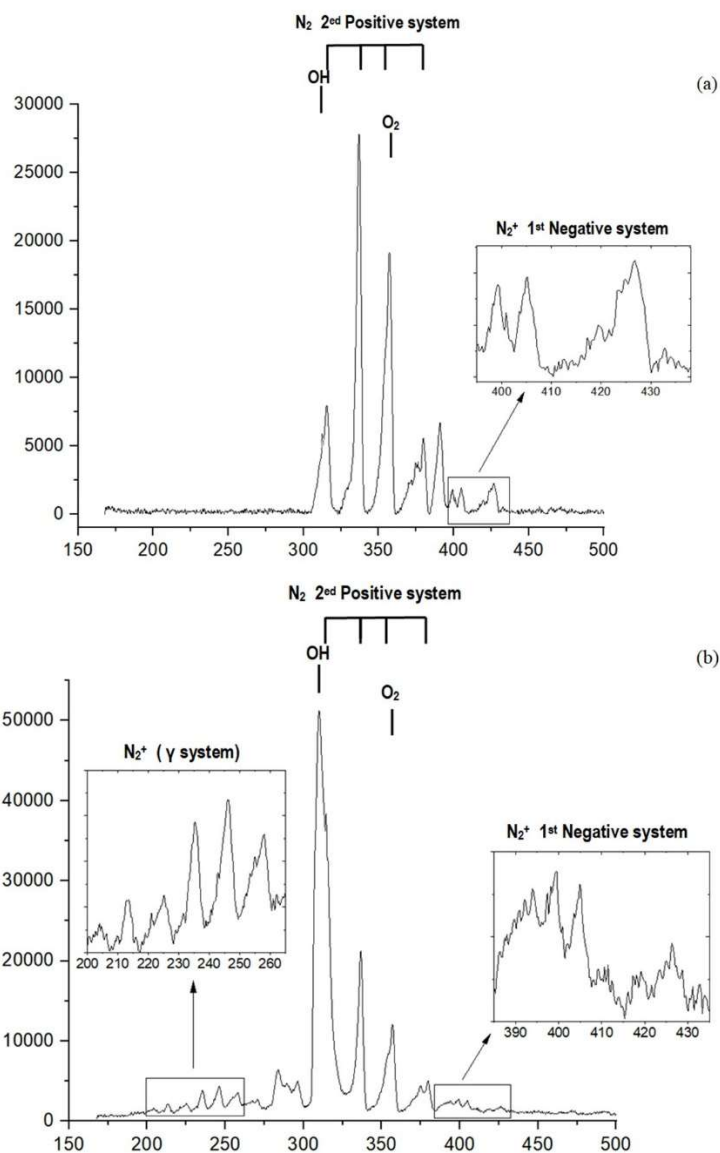

FIG. 8. Spectrum of plasmas formed by a) electrical discharge on the water surface and b) arc discharge in air

Table I Physicochemical properties of activated waters

|     | Control   | HRW      | PAW1       | PAW2      | HPAW1     | HPAW2      |
|-----|-----------|----------|------------|-----------|-----------|------------|
| pH  | 5.5±0.05  | 5.73±0.2 | 4.71±0.015 | 3.96±0.24 | 4.56±0.09 | 3.39±0.31  |
| EC  | 1±0.01    | 14±3.5   | 48±2       | 194±2     | 55±1      | 470±15.35  |
| TDS | 2±0.01    | 7±0.75   | 24±1       | 97±1      | 27±0.5    | 240±7.7    |

EC: Electrical Conductivity; TDS: Total Dissolved Solids

arc in water, due to the presence of many reactive oxygen species, the intensity of excited nitric oxide species and hydroxyl radical is the highest. Low-intensity (N2 - the initial negative system) is also observed at wavelengths ranging from 390 to 430 nm[44]. Oxygen ($O_2$) is also present in the discharge, as shown by the emission spectrum's peak at 356 nm[45]. All of these reactive species are short-lived and produce reactive species that are long-lived[8,40,43,46].

These excited $N_2$ molecules react with $O_2$ in the air and turn it into NOx. The dissolution of produced NOx in water leads to the formation of stable active nitrogen species (RNS) such as $NO_2^-$ and $NO_3^-$ ions (nitrogen and nitric acid). When water comes into direct contact with plasma, it heats up and vaporizes, causing the formation of plasma through the interaction of air and water molecules. This process produces more oxygenated and nitrogenated species, which can be seen in the plasma spectrum.

### C. Physicochemical Properties of water

The pH, electrical conductivity and total dissolved solids of the activated water were measured to determine their physicochemical properties. The concentration of hydrogen ions in the solution is measured by the pH. Table I shows a decrease in pH levels in PAWs, but injecting hydrogen gas leads to a slight increase in pH. The pH changes in PAWs acidification are consistent with other articles[9,12–14]. Hopkins' paper indicates that pH levels should rise in HRW[47].

The reaction of water with the species produced in the plasma causes PAW to become acidic. The PAW shall become more acid as the quantity of ROS and RNS delivered to the water increases. Changes in ionic species within PAWs were first identified by measuring their electrical conductivity. The increases in conductivity are due to the formation of ROS and RNS, when active water is prepared.

Hydrogens are the lightest gases and they react with most elements under normal conditions, whereas $H_2$ is not very reactive in its molecular state[48]. At the interface of plasma with water, hydrogen molecules are also decomposed in sample HPAW2. More reactions occur, and hydrogen radicals produce more of the secondary compounds. There is a growing possibility that more compounds will be dissolved in the water.

### D. Water Raman spectra

The Raman spectra of the water samples were collected using a Technooran (Microspectrophotometer, Ram-532-004) in the range of 134–4325 cm$^{-1}$ with an exposure time of 500 ms and three accumulations. Activated water's Raman spectrum is illustrated in FIG. 9. Using the reference data available in the literature, the Raman spectrum of water can be seen in FIG.9 lines (A) and (B), with vibration bands near 3400 and 3250 cm$^{-1}$ [49]. The Raman Q branch of hydrogen dissolved in water is observed at wavelengths ranging from 3900 cm$^{-1}$ to 4170 cm$^{-1}$[50]. In FIG. 9., line (c) represents a peak of the Q-branch.

Configuration of two hydrogen bonds and two covalent bonds between oxygen and hydrogens around oxygen atom is consistent with the tetrahedral-like structure of five water molecules[51]. As the impurities increase in the water, the four-hole configuration is prevented from forming, and the peak height is reduced. Alternatively, hydrogen molecules can contribute to this structure's formation and intensify its peak.

In addition to HRW, HPAW1 and HPAW2, hydrogen peak can also be observed in PAW1 and PAW2 due to reactions between plasma active species and liquids. Water peak intensity has decreased in PAW2 compared to PAW1 and HPAW2 compared to HPAW1 due to the increasing number of substances soluble in water. By forming a symmetrical configuration, hydrogen molecules in HPAW1/ HPAW2 have intensified the Raman Q-branch peak of water compared to PAW1/ PAW2.

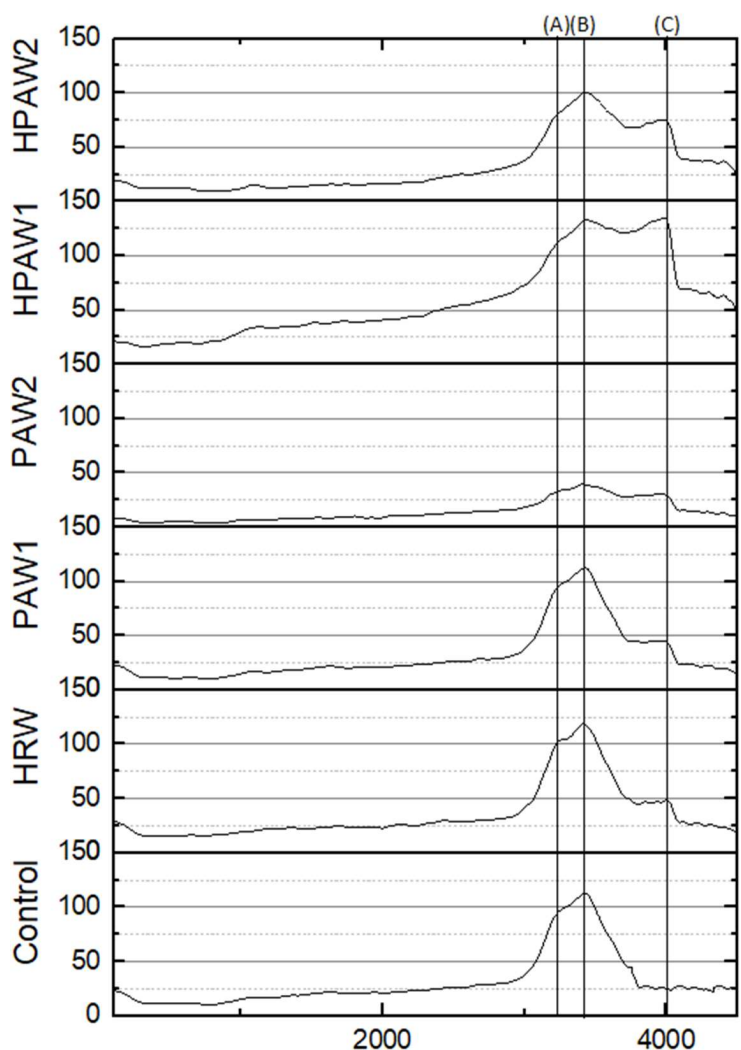

*FIG. 9.* The Raman spectra of activated waters

### E. Length and germination parameters

The measurement results of 125 lentil bean sprouts in each group were taken for length and germination parameters in five replicates on the third day (FIG.10). FIG.11 unequivocally displays the obtained results.

The acronym FGP represents the final percentage of germination, which serves as an indicator of the seed population's viability. The seeds from HPAW1 and HPAW2 had the highest FGP, with a reported percentage of 93.95%.

MGT, or Mean Germination Time, refers to the average time required for seeds to germinate, taking into account the timing of the majority of germination events. The lower the MGT, the faster a population of seeds has germinated[52]. The Control had the highest MGT, followed by HRW with a 10.5% decrease. PAW1 and HPAW1 both decreased by 13% and 12.7%, respectively. The lowest MGT value belonged to PAW2 and HPAW2 with a decrease of around 12.8% compared to the reference sample.

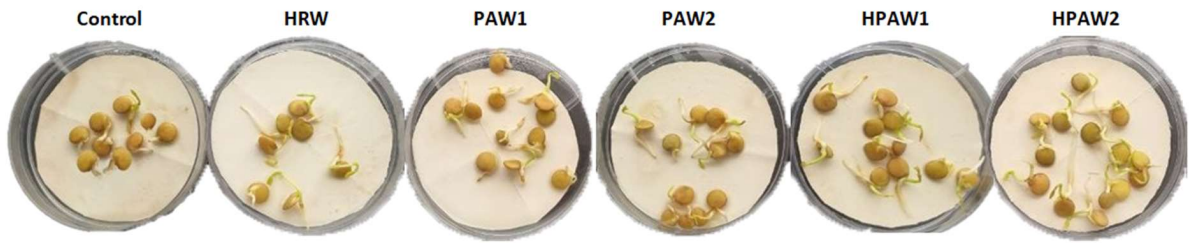

*FIG. 10. Lentil seeds treatment using PAW, HRW and control on the third day*

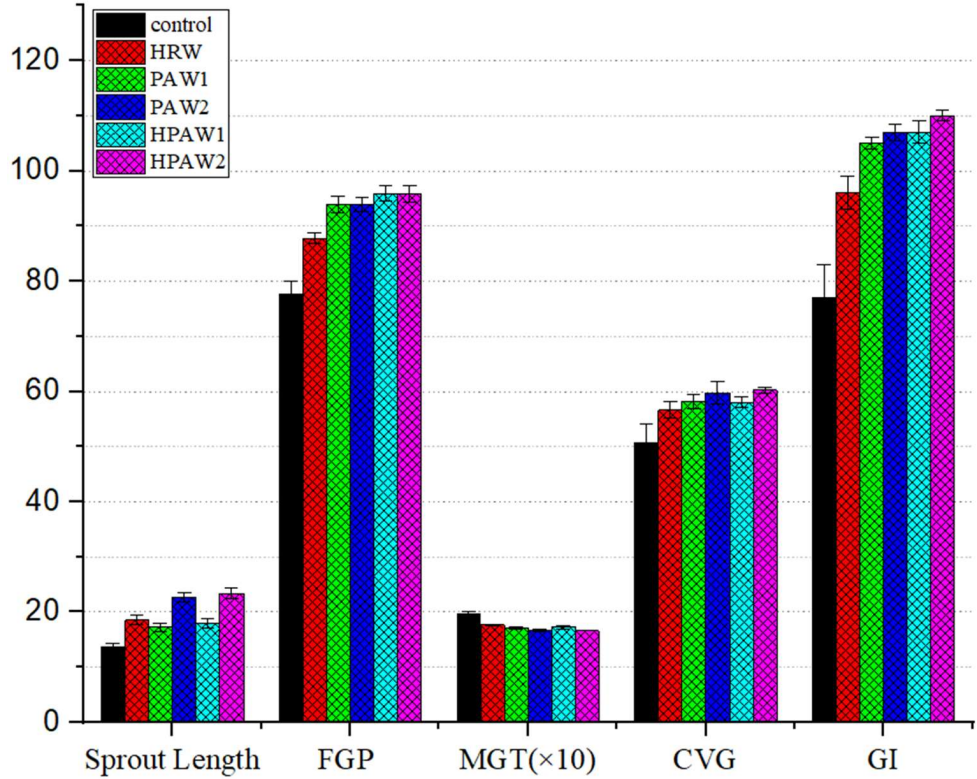

*FIG. 11. Change in agronomic traits of lentil seed after treatment with PAW, HRW and control*

The CVG (Coefficient of velocity of germination) solely measures the speed of germination based on the quantity of seeds and time, without placing emphasis on the final percentage[41]. It prioritizes the time taken to achieve the result and does not consider specific time details, only general averages[53]. In case of PAW2/HPAW2 and PAW1/HPAW1 samples, nearly identical CVG is reported. In comparison to the control sample, almost identical CVG ratios were reported for 1.2 and 3.4 samples with increases of 18.4% and 14.7% respectively. Compared to a reference sample, the CVG increased by 12.5% for sample HRW. PAW's RONS can enhance germination rates by activating the abscisic acid/gibberellic acid pathways, interrupting seed dormancy.

In the Germination Index (GI), seeds germinated on the first day are given more weight than those germinated later. The GI measures both the percentage and speed of germination, with a higher value indicating a higher percentage and rate[54]. The highest recorded value for GI is related to HPAW2, with a 42.9% increase compared to the reference. With an increase of 38.9%, it is followed by PAW2 and HPAW1. HRW and PAW1 both increased by 36.4 and 24.8 respectively. The GI is considered the most comprehensive measurement parameter, combining germination percentage, speed, spread, duration, and high/low events[41].

There is a progressive differentiation of organs and tissue in developing legume seeds. The maturation is governed by a signaling network composed of sugars, amino acids, and SnRK1 kinases. Changing levels of oxygen energy and the nutrient state activates maturation processes. Embryo cells become green and photosynthetically operative during the transition. Sucrose functions as a transport and nutrient sugar and as a signal molecule that activates storage-associated processes[55]. HRW plays a vital role in the conversion of starch to sugars[56], which are simple carbohydrates made of carbon, hydrogen, and oxygen[57]. The accumulation of soluble proteins not only aids water retention but also protects essential substances. This leads to maximizing plant growth and promoting good health[58].

## IV. Conclusion

In this study, different PAW reactors and hydrogen injection systems were used to produce active species in water. HRW-irrigated seeds have better parameters than control seeds, as evident from the results. The presence of RONS groups in PAW water stimulates seed metabolism and accelerates growth. The amount of compounds dissolved in water changes depending on the type of plasma system used. HPAW and PAW irrigation may result in a significantly higher yield in a shorter period of time, based on our findings. Future studies assessing the long-term effects of plasma and hydrogen treatment should examine the biological and chemical changes in the sprouts and their natural microbiota.

**Data Availability**

The data that support the findings of this study are available from the corresponding author upon reasonable request.